\documentclass[aps,prl,twocolumn,groupedaddress,showpacs,showkeys,amsfonts,amssymb,amsmath]{revtex4}
\usepackage{graphics}
\usepackage[colorlinks=true,linkcolor=blue,citecolor=blue,urlcolor=blue]{hyperref}

\begin{document}

%Title of paper
\title{THz Magneto-electric atomic rotations in the chiral compound Ba$_3$NbFe$_3$Si$_2$O$_{14}$.}

\author{L. Chaix}
\affiliation{Institut N\'eel, CNRS et Universit\'e Joseph Fourier, BP166, F-38042 Grenoble Cedex 9, France}
\altaffiliation{Institut Laue Langevin, 6 rue Jules Horowitz, BP 156, F-38042 Grenoble Cedex 9, France}

\author{ S. de Brion}
\email{sophie.debrion@grenoble.cnrs.fr}
\affiliation{Institut N\'eel, CNRS et Universit\'e Joseph Fourier, BP166, F-38042 Grenoble Cedex 9, France}

\author{F. L\'evy-Bertrand}
\affiliation{Institut N\'eel, CNRS et Universit\'e Joseph Fourier, BP166, F-38042 Grenoble Cedex 9, France}

\author{V. Simonet}
\affiliation{Institut N\'eel, CNRS et Universit\'e Joseph Fourier, BP166, F-38042 Grenoble Cedex 9, France}

\author{R. Ballou}
\affiliation{Institut N\'eel, CNRS et Universit\'e Joseph Fourier, BP166, F-38042 Grenoble Cedex 9, France}

\author{B. Canals}
\affiliation{Institut N\'eel, CNRS et Universit\'e Joseph Fourier, BP166, F-38042 Grenoble Cedex 9, France}

\author{P. Lejay}
\affiliation{Institut N\'eel, CNRS et Universit\'e Joseph Fourier, BP166, F-38042 Grenoble Cedex 9, France}

\author{J. B. Brubach, G. Creff, F. Willaert, and  P. Roy  }
\affiliation{Synchrotron SOLEIL, L'Orme des Merisiers Saint-Aubin, BP 48, F-91192 Gif-sur-Yvette Cedex, France}

\author{A. Cano}
\affiliation{European Synchrotron Radiation Facility, 6 rue Jules Horowitz, BP 220, 38043 Grenoble, France}
\date{\today}

\begin{abstract}
We have determined the terahertz spectrum of the chiral langasite Ba$_3$NbFe$_3$Si$_2$O$_{14}$ by means of synchrotron-radiation measurements. Two excitations are revealed that are shown to have a different nature. The first one, purely magnetic, is observed at low temperature in the magnetically ordered phase and is assigned to a magnon. The second one persits far into the paramagnetic phase and exhibits both an electric and a magnetic activity at slightly different energies. This magnetoelectric excitation is interpreted in terms of atomic rotations and requires a helical electric polarization.

\end{abstract}

\pacs{75.85.+t, 78.30.-j, 78.20.Bh}
%75.85.+t	Magnetoelectric effects, multiferroics (for multiferroics and magnetoelectric films, see 77.55.Nv)
%78.30.-j	Infrared and Raman spectra
%78.20.Bh	Theory, models, and numerical simulation
%75.25.Dk	Orbital, charge, and other orders, including coupling of these orders
%PACS 71.27.+a  Strongly correlated electron systems

\keywords{THz spectroscopy, magneto-electric excitations, langasite}

\maketitle

%\section{Introduction}

%%%%%%%%%%%%%%%%%%%%%%%%%%%%%%%%%%%%%%%%%%%%%%%%%
The electric-field control of spins and the converse magnetic-field control of electric dipoles inspire a number of hybrid technologies and motivates fundamental research on multiferroics and magnetoelectric materials \cite{BIB07}. These magnetoelectric couplings produce striking phenomena both at the static and at the dynamical level. A prominent example is the electric-charge dressing of magnons, resulting in the so-called electromagnons, that has been demonstrated experimentally in multiferroics \cite{PIM06,KIDA09}. This dressing enables the electric-field control of spin-waves and produces enhanced optical responses with potential applications in photonics and magnonics \cite{ROV10}. Here we report experimental evidence of the dual phenomenon, that are atomic vibrations dressed with currents, hence magnetically active, in the terahertz (THz) spectrum of the chiral compound Ba$_3$NbFe$_3$Si$_2$O$_{14}$. Based on symmetry arguments, we propose that this new type of magnetoelectric excitations is associated to structural rotations; their unexpected magnetoelectric activity is triggered by a spontaneous helical polarization in the paramagnetic phase of the system.

\begin{figure}[b]
\resizebox{8.6cm}{!}{\includegraphics{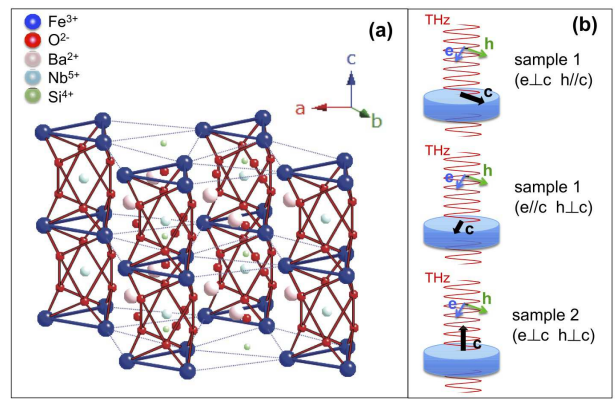}}
\caption{(a) Ba$_3$NbFe$_3$Si$_2$O$_{14}$ langasite crystallographic structure. (b) Relative orientation of the electric {\bf e} and magnetic {\bf h} fields of the THz wave with respect to the crystal {\bf c}-axis for samples 1 and 2.}
\label{fig1}
\end{figure}

When chirality meets magnetic order, unusual magnetic and electric properties often appear encouraging the search for novel multiferroic and magnetoelectric phenomena. The highest magnetically-induced polarization observed so far has been reported in CaMn$_7$O$_{12}$ \cite{Zhang11} where, interestingly, axial distortions of the lattice play a key role in generating ferroelectricity out of the chiral magnetic structure of the system \cite{Johnson12,Perks12}. We hereafter focus on the Ba$_3$NbFe$_3$Si$_2$O$_{14}$  langasite, which crystallizes within the non-centrosymmetric space group P321 \cite{MIL98} thus displaying
\emph{structural} chirality (see Fig.~\ref{fig1}). The remarkable magnetic properties of this system are due to the Fe$^{3+}$ ions. These form triangles stacked along the {\bf c}-axis and compose a triangular lattice in the ({\bf a}, {\bf b}) plane (see Fig.~\ref{fig1}).
The magnetic exchanges along the {\bf c}-axis are mediated by twisted Fe-O-O-Fe paths. Below the N\'eel temperature T$_N$~=~27~K, the system develops a chiral magnetic order with a 120$^{\circ}$  spin structure within the
\textcolor{blue}{Fe-}
triangles that rotates along the {\bf c}-axis and forms magnetic helices \cite{MAR08}.
In addition, below T$_N$, the system shows a static magnetoelectric effect \cite{MAR10} and chiral spin-wave excitations \cite{LOI11,JEN11,SIM12}.
A weak electric polarization along the {\bf c}-axis has also been reported \cite{ZHO09}, although our own measurements show no uniform polarization either along the {\bf c}-axis nor perpendicular to it.

We have studied Ba$_3$NbFe$_3$Si$_2$O$_{14}$ by means of synchrotron-radiation measurements in the THz regime. This technique is particularly well suited to investigate magnetoelectric excitations \cite{KIM12} since (i) the THz energy range coincides with that of magnons and phonons and (ii)
the THz-wave delivered by the synchrotron radiation (98 \% linearly polarized) carries both an electric and a magnetic field, so that rotating the crystal with respect to the wave polarization direction allows the unambiguous determination of the excitation fields. We have used these selection rules and the temperature dependence of the excitations to elucidate their nature and the mechanisms at play in the dynamics. This is further ascertained by comparison with inelastic neutron scattering measurements.

%\section{Experimental details}
Two plaquettes of $mm$ size were cut from the same single crystal \cite{MAR08} with a thickness of 450~$\mu m$ and 300~$\mu m$; the {\bf c}-axis was oriented either in plane or out of plane in order to probe all possible geometries of the THz electric and magnetic fields as regards the crystal {\bf c}-axis: ({\bf e//c}~~{\bf h$\perp$c}) and ({\bf e$\perp$c}~~{\bf h//c}) for one crystal, ({\bf e$\perp$c}~~{\bf h$\perp$c}) for the other crystal (see Fig.~\ref{fig1}). THz absorption spectra were obtained in the energy range 8-60~cm$^{-1}$ (0.24-1.8~THz) by measuring the transmission at the AILES beamline of Synchrotron SOLEIL \cite{SOL}. A Bruker IFS125 interferometer equipped with a pulse tube cryostat was used, combined with a Helium pumped bolometer. The 10-60~cm$^{-1}$ energy range was explored at a resolution of 0.5~cm$^{-1}$ using a 6~$\mu m$ thick silicon-mylar multilayered beamsplitter (BMS). The 8-18~cm$^{-1}$ energy range was explored with more sensitivity by using a 125~$\mu m$ thick mylar BMS.

%\section{Results}

The THz spectra measured at various temperatures from 300 K to 6 K (for more details, see Supplemental Material) reveal an absorption roughly three times larger for {\bf e//c} than for {\bf e$\perp$c}, meaning that the dielectric response of the material is considerably stronger for this direction of the THz electric field. This additional absorption arises from (multi-)phonons excited by {\bf e//c}, that contribute as a broad background centered at 40~cm$^{-1}$ growing with increasing temperature, as expected for thermally populated phonons.

\begin{figure}
\resizebox{8.6cm}{!}{\includegraphics{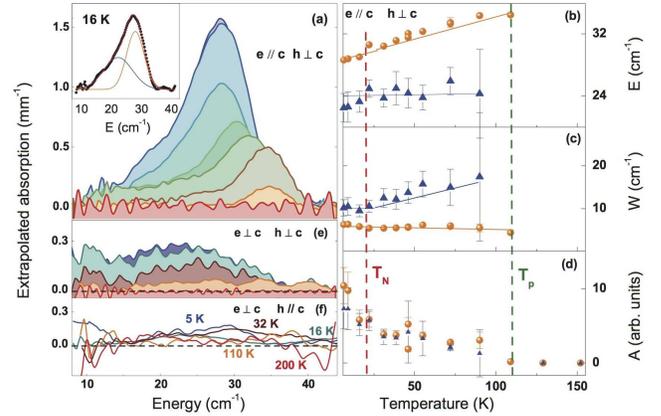}}
\caption{\textbf{THz excitations probed with the 6~$\mathbf{\mu}$m BMS.} (a)~Extrapolated absorption spectra for ({\bf e//c}~~{\bf h$\perp$c}) at different temperatures. Insert~: gaussian  fits. (b)(c)(d)~Temperature dependance of the fits position, width and area. (e)(f)~Same as (a) for the other THz polarizations.}
\label{fig2}
\end{figure}

 Since the magnetic contribution is weak compared to the instrument and phononic background, two different methods were used
to enhance the magnetoelectric response of the samples. For the 6~$\mu m$ BMS measurements, we used the same procedure as in \cite{DEB07}: we estimated the extrapolated absorption at a given temperature T$_{0}$, $\alpha$(T$_{0}$), as the common signal to the  $\alpha$(T$_{0}$)-$\alpha$(T) curves for all measurements temperatures T.  A linear background was then subtracted. On the other hand, for the 125~$\mu m$  BMS measurements, the sample at 45 K was used as a common reference. Interference patterns arising from coherent reflections on the disk surfaces, when present, were removed using a band block FFT filter. The extrapolated absorption for all the different combinations of the THz electric field \textbf{e} and magnetic field  \textbf{h} alignment is displayed on Fig.~\ref{fig2}.

The most stiking feature is visible for ({\bf e//c}~~{\bf h$\perp$c}): a large asymmetric absorption is clearly observed that results from two gaussian contributions (see insert of Fig.~\ref{fig2}a).  At 16~K for instance, the two peaks are found at 23 and 29~cm$^{-1}$ with a respective width of 10 and 5~cm$^{-1}$. Rotating the electric field perpendicular to {\bf c} leaves with a single broad absorption centered around 25~cm$^{-1}$ (see Fig.~\ref{fig2}e). When an additional rotation is performed and the magnetic field lies parallel to {\bf c} no clear absorption is observed (see Fig.~\ref{fig2}f). These measurements evidence two excitations quite close in energy, E1 and E2, that persist in the paramagnetic phase up to 100~K, that is almost four times higher than T$_N$ (see Fig.~\ref{fig2}). This is a strong indication that they have a common origin which is not related to the long range magnetic order although they lie in the magnon energy range. It leads us to introduce a new characteristic temperature in the system, T$_{p}~\simeq ~$100~K.

\begin{figure}
\resizebox{8.6cm}{!}{\includegraphics{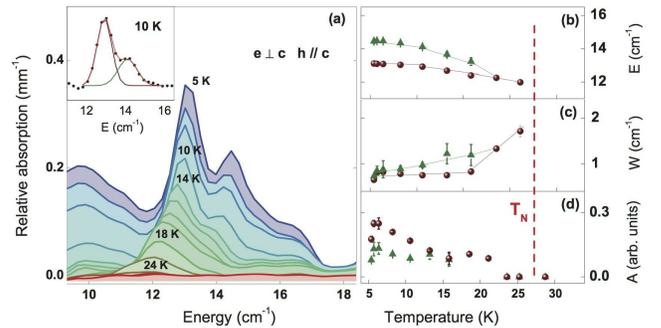}}
\caption{\textbf{THz excitations probed with the 125~$\mathbf{\mu}$ BMS.} (a)~Relative absorption spectra, abs(T)-abs(45~K) ({\bf e$\perp$c}~~{\bf h//c}) at different temperatures. Insert~: gaussian  fits. (b)(c)(d)~Temperature dependance of the fits position, width and area.}
\label{fig3}
\end{figure}
The overall behavior of these excitations presents some discrepancies: the energy of excitation E1 remains constant while for E2 it increases at high temperature. The life time (inverse of the absorption peak width) starts to decrease above T$_N$ for excitation E1 while it remains constant for E2. The most remarkable difference concerns their selection rules: E1 is activated by the magnetic field {\bf h$\perp$c}, E2  by the electric field {\bf e$\parallel$c}  [see Fig 2 panels (a), (e) and (f)]. Note that both E1 and E2 are clearly seen in panel (a), E1 is still visible in panel (e), and no feature is detectable in panel (f).

Another weaker and even sharper excitation (E3) is found at lower energy for ({\bf e$\perp$c}~~{\bf h//c}), using the more sensitive 125 $\mu$m BMS,  resulting from two close gaussian contributions. At 10 K, they are observed at 12.9~cm$^{-1}$ and 14.1~cm$^{-1}$, both with a width of 0.8~cm$^{-1}$ (see Fig.~\ref{fig3}a). This excitation shifts to lower energy and broadens as the temperature is increased to finally disappears at 24~K, just below T$_N$. Contrary to the previous excitations, the temperature dependence of this excitation is characteristics of a magnon.

%neutrons

\begin{figure}[tb]
\resizebox{8.6cm}{!}{\includegraphics{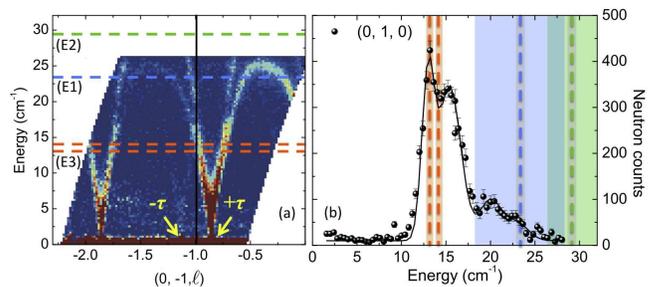}}
\caption{\textbf{Magnons probed by inelastic neutron scattering}. (a)  Magnons dispersion along the (0,-1,-$\ell$) direction in the reciprocal space. The black line indicates the Brillouin zone center. (b) Energy scan at the reciprocal lattice zone center (0,1,0). The colored dash lines give the energy position of excitations E1, E2 and E3 seen in THz measurements while the colored rectangles indicate their width. }

\label{fig4}
\end{figure}

This was confirmed from the comparison with the magnon spectra obtained through inelastic neutron scattering \cite{LOI11,STO11,JEN11,SIM12}. The THz measurements probe the first Brillouin zone center (0,0,0) where no neutron measurements are available. We therefore compared them to neutron spectra in other zone centers. Selected  measurements performed on the time-of-flight IN5 and on the triple-axis IN12 spectrometers at the Institut Laue Langevin are shown in Fig.~\ref{fig4}. Two main branches emerge from the magnetic satellites at $+\tau$ with $\tau$=1/7 \cite{NOTE1}. The lower energy branch is associated with the dynamical correlation function of spin components along the {\bf c}-axis. The upper energy branch is associated with those in the triangle plane. An intense signal is observed around 13~cm$^{-1}$ at the zone center, mainly from the lower branch (see Fig. \ref{fig4}b). The tail of this signal and the less intense neutron signal up to 25~cm$^{-1}$ are attributed to the upper branch. The first neutron peak at 13~cm$^{-1}$ perfectly agrees with the weak excitation E3 observed in the THz measurements in terms of energy and spin component involved (along the {\bf c}-axis, hence excitable by {\bf h//c}). The other magnonic signals (involving spin components in the ({\bf a},{\bf b}) plane) are not clearly distinguished in the THz measurements for {\bf h$\perp$c}, presumably because the 10 times stronger excitations E1 and E2 mask everything (see Fig.~\ref{fig2}a and Fig.~\ref{fig2}e). These remarkable larger THz-excitations E1 and E2 are clearly absent from the neutron data of Fig.~\ref{fig4} and also from references \cite{LOI11,STO11,JEN11}.

In multiferroics, magnons can be electro-active at  particular points of the Brillouin zone (related  to the spiral wave vector or at the Brillouin zone edges) \cite{PIM06, KIDA09, KBN07}
and these so called electromagnons can persist in the non-multiferroic, yet magnetically ordered phases of these compounds \cite{PIM06, KIDA09, CAN09}. In contrast, for the langasite compound, the THz excitations E1 and E2 persist up to $100~\text{K} \approx4 T_{N}$ when no well-defined magnetic excitation exists anymore \cite{SIM}. Thus, whereas the THz-excitation (E3) at 13~cm$^{-1}$ can be unambiguously identified as a conventional magnon, the magnonic origin of the THz-excitations E1 and E2 can be discarded.

These excitations are more likely connected to the lattice \cite{NOTE2}, although their relatively low energies and the unusual behavior described above is different from that of standard phonons. The natural candidates to explain these features are axial distortions of the lattice, whose role has been recently highlighted in this class of materials \cite{FIE11,Johnson12,Hearmon12}. These are rotations $R_z$ about the {\bf c}-axis whose interaction with the electric and magnetic fields of the external radiation is described by the effective Lagrangian:
\begin{align}
%L_\text{int} =  {\alpha \over \omega _{P_ z}^ 2 } [ E_z  + i g  (\dot {\mathbf H}_\perp\times {\mathbf P}_\perp)_z  ] R_z.
L_\text{int} = {\alpha \over \omega _{P_ z}^ 2 } [ e_z  + g
({\mathbf P}_\perp  \times {\dot {\mathbf h}}_\perp )_z  ] R_z.
\label{Lint}\end{align}
This interaction results from the linear coupling between rotations and polar displacements allowed by the P321 space-group symmetry of our system (see Supplemental Material for details on the derivation of this Lagrangian). Here $\alpha$ represents the strength of this coupling, ${\mathbf P}$ is the electric polarization, with $\omega_{P_z}$ being its characteristic frequency (or energy), and $g$ is a gyromagnetic factor.

The excitation of $R_z$ by the electric-field component of the radiation observed in our THz measurements is described by the first term in Eq. \eqref{Lint}. This follows immediately from the aforementioned linear coupling between $R_z$ and $P_z$.
The additional response of $R_z$ to the magnetic field is due to the second term in Eq. \eqref{Lint}. This response, in contrast, is far more subtle. It requires a finite, static polarization perpendicular to the {\bf c}-axis, $ P_\perp \not = 0$, to be operative. At room temperature, no perpendicular polarization is allowed by the symmetry. However, at low temperatures, recent neutron scattering results are not compatible with a perfect 120$^{\circ}$ spin structure, implying the loss of the three-fold symmetry axis \cite{SIM}. This $P321\to C2$ symmetry breaking implies the appearance of a spontaneous polarization perpendicular to the {\bf c}-axis $\mathbf P_\perp$ that, in turn, activates the coupling of $R_z$ with the magnetic field in Eq. \eqref{Lint}. This is in perfect agreement with the gradual appearance of the magneto-active excitation E1 that we observe at 24 cm$^{-1}$ below 100~K (see Fig.~\ref{fig2}).
The concomitant appearance of the electro-active excitation E2 at 100~K can be explained as due to the increase of the overall effective coupling  ${\alpha / \omega_{P_z} ^2 }$ at the onset of the new phase. This increase is likely a by-product of the softening of $\omega_{P_z}$ due to its dependence on $P_\perp$. In fact, we observe a similar softening in the energy of the electro-active excitation itself that is in tune with this hypothesis (see Fig.~\ref{fig2}b).

\begin{figure}[t!]
\resizebox{8.6cm}{!}{\includegraphics{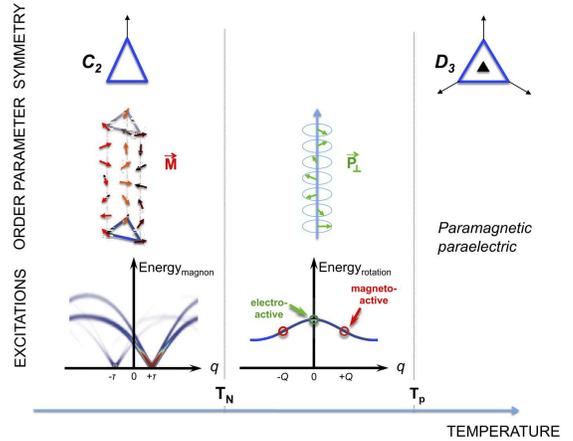}}
\caption{\textbf{Temperature evolution of the electric and magnetic phases.} From top to bottom: symmetry, order parameter, and corresponding excitations. From right to left (decreasing temperature): T~$>$~T$_p$ paramagnetic and paraelectric state, T$_p>$~T~$>$~T$_N$ helicoidal state of polarization, T$_N>~$T additional helicoidal magnetic state. The P321 to C2 symmetry breaking is schematized by the loss of the $C_3$ axis in the Fe$^{3+}$ triangles (from equilateral to isosceles). The curvature of the dispersion curve $E_\text{rot}(q)$ is deduced from the relative energies of the magnetic and electric active peaks $E1$ and $E2$.}
\label{fig6}
\end{figure}

Finally, we note that in our spectra the energy of the electro-active and magneto-active excitations are slightly different.
This difference suggests that the spontaneous polarization that appears below 100 K is modulated with a finite wavevector $Q$. Thus, according to Eq. \eqref{Lint}, the electric field couples to $q=0$ uniform rotations while the magnetic field does to $q=\pm Q$ rotations whose resonant energy is lower (see Fig. 5). The detailed Ginzburg-Landau analysis of the $P321\to C2$ transition confirms this possibility. The perpendicular polarization $ P_\perp$ can be seen as the order parameter driving the $P321\to C2$ transition. The Ginzburg-Landau free energy then can be written in terms of  $(P_x,P_y)=(\rho \cos\theta, \rho \sin\theta )$
as:
\begin{align}
F =& {a\over 2} \rho^ 2 + \gamma \rho^3 \cos 3 \theta + {b\over 4}\rho^ 4
\nonumber \\&
- \lambda \rho^2 (\partial_z \theta) + {c \over 2} \big[ (\nabla \rho)^2 + \rho^2 (\nabla \theta)^ 2  \big].
\label{F}
\end{align}
Here, as customary, $a$ is assumed to be the only coefficient that depends on the temperature and changes sign as $a = a'(T - T_0)$, where $T_0$ can be seen as the nominal transition temperature. If the space variations of the order parameter are ignored, then the cubic term in the free energy implies rather a first-order transition at $T_0 ' = T_0 + 2\gamma^2/(a' b)$.
The presence of the Lifshitz invariant $- \lambda \rho^2 (\partial_z \theta)$, however, leads to a second-order transition at higher temperature, which sets up helical spontaneous polarization $(P_x, P_y) = (P_0\cos Q z , P_0\sin Q z )$ at $T_P = T_0 + \lambda^2/(a' c)$ (see Fig.~\ref{fig6}). Here $P_0^2 = |T-T_P|/(a'b)$ and $Q = \lambda /c $. This will actually be the case if $2 \gamma^2 /b < \lambda^2/c  $ (that is, if the expected discontinuity at the first-order transition is relatively weak). This type of helical polarization has been discussed theoretically in the context of incommensurate phases, although no experimental realization has been reported so far \cite{LEV86}. Here $T_P$ is different from the direct $P321\to C2$ transition temperature, which is avoided due to the formation of the intermediate helical state. The $C2$ symmetry is obtained when the helix becomes commensurate. This is expected to take place at a lower temperature, which may well coincide with the magnetic transition temperature.

The essential ingredients that give rise to the magnetoelectric excitations in our system are also present in the newly reported ferroaxial multiferroic CaMn$_7$O$_{12}$ and related compounds \cite{Zhang11,Johnson12,Perks12,Hearmon12}. There, the linear coupling between the electric polarization and macroscopic structural rotations \cite{Johnson12,Hearmon12}, is preceded by a structural helix due to the incommensurate modulation of the atomic bonds \cite{Perks12}, in CaMn$_7$O$_{12}$ at least. However a chiral magnetic order is necessary to activate the polar-axial coupling via the trilinear invariant $\sigma P_z R_z$, where $\sigma$ represents the chirality of the magnetic structure. In the langasites, the lattice itself is chiral, so that the polar-axial coupling is ``switched on'' already in the paramagnetic phase.
Thus, while structural rotations in both these systems can be expected to have similar magnetoelectric responses, they can reveal different key aspects associated to structural vs. magnetic chirality.

%\section{Conclusion}
In summary, our THz investigation of the langasite dynamical properties has revealed a new kind of excitation, associated to atomic vibrations, which is both electric and magnetic active, at slighly different energies. These findings demonstrate that not only magnons but also atomic vibrations can acquire a magnetoelectric character, thus opening new routes to carry and process information.
This striking magnetoelectric activity has been related to the bilinear coupling between the atomic rotations allowed by the chiral symmetry of our system and a perpendicular electric polarization. The predicted helical polar order may itself exhibit original properties such as dielectric chirality and sets off new magnetoelectric functionalities that can be common among symmetry-equivalent systems.

\begin{acknowledgments}
We are very grateful to J. Ollivier, M. Enderle and P. Steffens for their help during the neutron experiments on IN5 and IN12. We thanks P. Bordet for discussions on the langasite structure.  We acknowledge J. Debray, J. Balay and A. Hadj-Azzem for the crystals preparation.
\end{acknowledgments}

%%%%%%% References


\begin{thebibliography}
\bibitem{NOTE1} The $-\tau$ satellites in the presented Brillouin zones are much weaker as a consequence of the structural chirality \cite{LOI11}.
\bibitem{NOTE2} Note that the neutron spectra measured only within the first and second Brillouin zones can scarcely probe phonons.
\end{thebibliography}
\end{document}

%Title of paper
\title{THz Magneto-electric atomic rotations in the chiral compound Ba$_3$NbFe$_3$Si$_2$O$_{14}$.}

\author{L. Chaix}
\affiliation{Institut N\'eel, CNRS et Universit\'e Joseph Fourier, BP166, F-38042 Grenoble Cedex 9, France}
\altaffiliation{Institut Laue Langevin, 6 rue Jules Horowitz, BP 156, F-38042 Grenoble Cedex 9, France}

\author{ S. de Brion}
\email{sophie.debrion@grenoble.cnrs.fr}
\affiliation{Institut N\'eel, CNRS et Universit\'e Joseph Fourier, BP166, F-38042 Grenoble Cedex 9, France}

\author{F. L\'evy-Bertrand}
\affiliation{Institut N\'eel, CNRS et Universit\'e Joseph Fourier, BP166, F-38042 Grenoble Cedex 9, France}

\author{V. Simonet}
\affiliation{Institut N\'eel, CNRS et Universit\'e Joseph Fourier, BP166, F-38042 Grenoble Cedex 9, France}

\author{R. Ballou}
\affiliation{Institut N\'eel, CNRS et Universit\'e Joseph Fourier, BP166, F-38042 Grenoble Cedex 9, France}

\author{B. Canals}
\affiliation{Institut N\'eel, CNRS et Universit\'e Joseph Fourier, BP166, F-38042 Grenoble Cedex 9, France}

\author{P. Lejay}
\affiliation{Institut N\'eel, CNRS et Universit\'e Joseph Fourier, BP166, F-38042 Grenoble Cedex 9, France}

\author{J. B. Brubach, G. Creff, F. Willaert, and  P. Roy  }
\affiliation{Synchrotron SOLEIL, L'Orme des Merisiers Saint-Aubin, BP 48, F-91192 Gif-sur-Yvette Cedex, France}

\author{A. Cano}
\affiliation{European Synchrotron Radiation Facility, 6 rue Jules Horowitz, BP 220, 38043 Grenoble, France}
\date{\today}

\maketitle
\onecolumngrid
\section{Supplemental material}
\subsection{THz absorption spectra in the two Ba$_3$NbFe$_3$Si$_2$O$_{14}$ crystals}

\begin{figure}[h]
\resizebox{8.6cm}{!}{\includegraphics{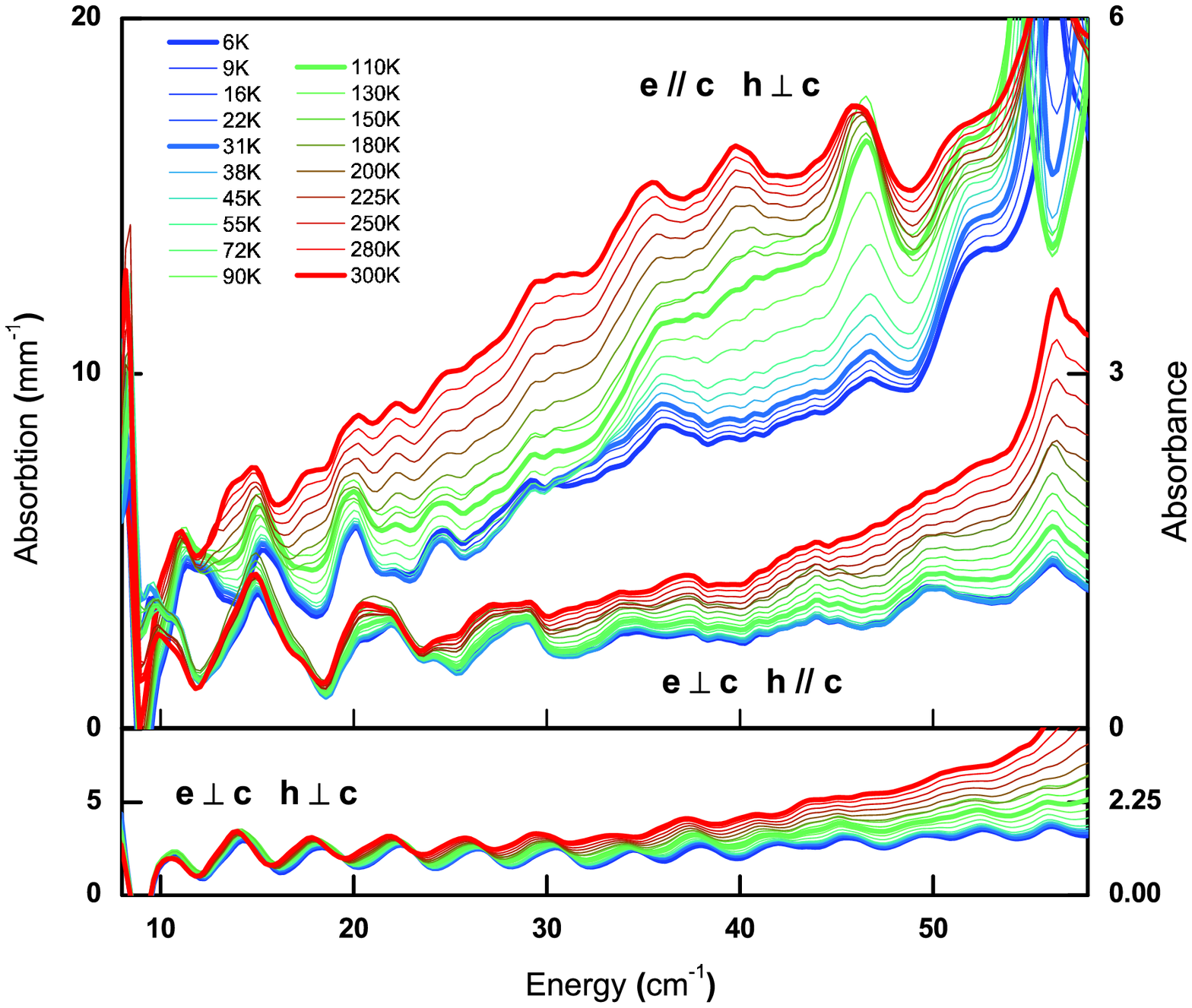}}
\caption{\textbf{Absorption spectra in Ba$_3$NbFe$_3$Si$_2$O$_{14}$ measured with the 6~$\mathbf{\mu}$m BMS}. The temperature was varied from 6 K to 300 K. Three different polarizations of the THz electromagnetic field as regards the crystal {\bf c}-axis were measured. Upper panel: measurements on sample 1, ({\bf e//c}~~{\bf h$\perp$c}) and ({\bf e$\perp$c}~~{\bf h//c}). Lower panel: measurements on sample 2 ({\bf e$\perp$c}~~{\bf h$\perp$c}). }
\label{fig}
\end{figure}

Figure~\ref{fig} presents the absorption/absorbance spectra (left/right scales) measured with the 6~$\mathbf{\mu}$m BMS on two different samples. The absolute absorbances, $\alpha d$, with $\alpha$ the absorption and $d$ the travelled distance through the sample, were determined by measuring the transmission through a $2~mm$ diaphragm as a reference and the sample transmission through that same diaphragm. Interference pattern from the sample surfaces is present for sample 2 (lower panel). For further analysis, this interference pattern was removed using a band block FT filter. The larger oscillations observed below 30~cm$^{-1}$ for sample 1 (upper panel) are instrumental. For further analysis, they were removed using the sample at a different temperature as a reference. The stronger absorption observed for {\bf e//c} is attributed to a (multi-)phonon band.

\subsection{Interaction between $R_z$ and the external fields in Eq.1}

In isotropic or cubic systems, the external field conjugated to rotations is a torque. However, in systems with reduced symmetry, the linear coupling between rotations $\mathbf R$ and linear displacements can be allowed. This is precisely our case, since our system initially has $D_3$ point-group symmetry which is further reduced to $C_2$.
This means that $\mathbf R$ in our system is not a pure rotation anymore, so that the standard forces acting on the above displacements also act on $\mathbf R$.
Polar displacements implying a relative motion of charges are among the displacements linearly coupled to $\mathbf R$.
Thus, the electromagnetic radiation is coupled to rotations via this motion of charges, which gives rise to the unusual electromagnetic activity of the atomic rotations in our system.

More specifically, the charges feel the electromagnetic radiation through the Lorentz force: ${\mathbf f} =
-{e \over c} {\partial \mathbf A \over \partial t}  - e \nabla \phi + {e \over c}\dot {\mathbf r} \times (\nabla \times {\mathbf A})
$, where ${\mathbf r}$ represents the position of the charge $e$, while $\phi$ and ${\mathbf A}$ are the scalar and vector potentials of the radiation respectively \cite{LAN75}. In the Weyl gauge ($\phi = 0$), the vector potential of a monochromatic plane wave is ${\mathbf A}(\mathbf r , t) = {\mathbf a}_0 e^{i ({\mathbf k}\cdot {\mathbf r} - \omega t)} \simeq  (1 + i {\mathbf k}\cdot {\mathbf r}) {\mathbf A}_0(t)$ in the long-wavelength limit.
Here $k = n \omega/ c$ and ${\mathbf A}_0(t)= {\mathbf a}_0 e^{ - i\omega t}$, with $n$ being  the refractive index of the medium and ${\mathbf a}_0 $ a constant complex vector.
In this limit, the first term in the vector potential generates the electric field of the wave ${\mathbf E} = i (k/n) {\mathbf A}_0$, while the magnetic field ${\mathbf H } = i {\mathbf k} \times {\mathbf A}_0$ is obtained from the second one. Keeping this in mind, the vector potential can be rewritten as ${\mathbf A} \simeq  {n \over i k} \mathbf E + \mathbf H \times \mathbf r + n {\mathbf k \over k}(\mathbf E \cdot \mathbf r)
$.
Accordingly, the Lorentz force becomes ${\mathbf f} \simeq e {\mathbf E} + {e  \over c} {\mathbf r}\times \dot {\mathbf H}
+i e {\mathbf k} ( {\mathbf E}\cdot{\mathbf r})
+ {e \over c}\dot {\mathbf r} \times {\mathbf H} $.
We note that the second and the last contribution to this force are sort of dual contributions. Interestingly, the last term was invoked by Dzyaloshinskii and Mills in order to explain the unexpected paramagnetic behavior observed in nonmagnetic ferroelectrics \cite{DZY09}.
In our case, the second term is the key one for the physical interpretation of the absorption spectrum of our system. In fact, in our measurements, the last two contributions to the Lorentz force produce subdominant effects and therefore can be ignored. Furthermore, this force acts directly on the charge distribution of the system, and therefore on its electric polarization $\mathbf P = \sum  e_i {\mathbf r} _i$. But since this polarization is linearly coupled to atomic rotations in our system, this gives rise to an effective force acting on the rotations as explained above. The force acting the rotations $R_z$ about the $z$-axis, in particular, can be written as
\begin{align}
f_\text{eff} = {\alpha \over \omega _{P_ z}^ 2 } E_z  +g
{\alpha \over \omega _{P_ z}^ 2 }
({\mathbf P}_\perp\times \dot {\mathbf H}_\perp)_z .
\label{feff}\end{align}
Here $\alpha$ represents the strength of the coupling between $R_z$ and $P_z$,  $\omega_{P_z}$ is the frequency (or stiffness) associated to $P_z$, and $g$ is a gyromagnetic factor. This force can be obtained as $f_\text{eff} = {\delta L_\text{int}\over \delta R_z}$, where $L_\text{int}$ is the Lagrangian in Eq. (1) of the main text.

If ${\mathbf P}_\perp$ is uniform, both electric and magnetic-field components of the THz radiation %light in the long-wavelength limit
will couple to $q=0$ rotations due to momentum conservation. However, if ${\mathbf P}_\perp$ is modulated with a finite wavevector $Q$, then $L_\text{int}$ becomes
\begin{align}
L_\text{int} = \underset{k\to 0}{\lim}
\Big[ \alpha_1 E_z (k) R_z (-k)
+\alpha_2 \big({\mathbf P}_\perp (-Q) \times \dot {\mathbf H}_\perp (k)\big)_z R_z(Q - k) + (Q\to - Q )\Big]
\label{feff}\end{align}
in Fourier space, where $\alpha _1 = {\alpha / \omega _{P_ z}^ 2(0) }$ and $\alpha_2 = {g \alpha / \omega _{P_ z}^ 2(Q )}$. In this case, the electric field couples to $q=0$ rotations while the magnetic field does to $q= \pm Q$ ones, thus probing different resonant energies.
%%%%%%% References